\documentclass[%
 preprint,
 amsmath,amssymb,
 aps,
pre,
]{revtex4-1}

\usepackage{graphicx}
\usepackage{dcolumn}
\usepackage{bm}
\usepackage{color}
\usepackage{hyperref}

\newcommand{\yh}[1]{\textcolor{black}{#1}}

\begin{document}


\title{Detection and tracking of chemical trails by local sensory systems}

\author{Yangyang Huang$^1$, Jeannette Yen$^2$ and Eva Kanso$^1$}\thanks{Kanso@usc.edu}
 \affiliation{${}^1$ Aerospace and Mechanical Engineering Department, \\ 
 University of Southern California, Los Angeles, CA 90089}
 \affiliation{${}^2$ School of Biology, Center for Biologically-Inspired Design, \\Georgia Institute of Technology, Atlanta, GA 30332}
\date{\today}


\begin{abstract}                       
Many aquatic organisms  exhibit remarkable abilities to detect and track chemical signals when foraging, mating and escaping.  For example, the male copepod { \em T. longicornis}  identifies the female in the open ocean by following its chemically-flavored trail.  
Here, we develop a mathematical framework in which a local sensory system is able to detect the local concentration field and adjust its orientation accordingly.  We show that this system is able to detect and track chemical trails without knowing the trail's global or relative position. 
\end{abstract}
\maketitle


\section{Introduction}

The response to olfactory signals and pheromones plays an important role in a variety of biological behaviors~\cite{Partridg1993,Vickers2000,Zimmer2000} such as homing by the Pacific salmon~\cite{Hasler2012}, foraging by seabirds~\cite{Nevitt2000}, lobsters \cite{Basil1994,Devine1982} and blue crabs \cite{Weissburg1994}, and mate-seeking and foraging by zooplanktons and insects \cite{Carde1996,Carde1997}. These dissimilar organisms and behaviors share similar mechanisms of sensing and responding to chemical signals~\cite{Vickers2000}. The underlying mechanisms could be applied or adapted to design artificial devices for purposes such as source detecting in various environments, see examples in~\cite{Grasso2001,Pyk2006,Nakatsuka2006,Dhariwal2004}. 

\begin{figure}[!h]
\includegraphics[width=0.95\linewidth]{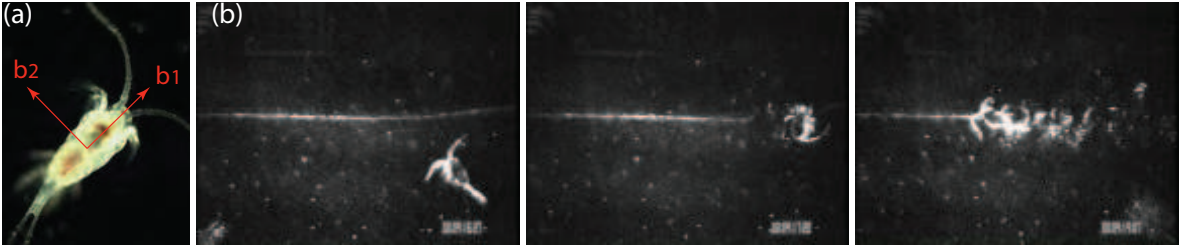}
\caption{(a) Copepod {\em Temora longicornis} ($\sim$ 1 mm in length). Sensing of chemical signals  is mediated by a distribution of small mechanoreceptive organs on its antennae. (b) Trail tracking  by copepod: sequences showing the progression of the copepod {\em T. longicornis} while navigating a chemically-flavored laminar trail mimic. The copepod first follows the trail in the direction away from the source then corrects its heading direction and traces the trail in the direction of increasing chemical signal.
 The trail mimic is created by releasing fluid via syringe pump (0.01 mL/min) and small bore tubing (1mm). Dextran, a large molecular weight, was added to increase refractive index of the trail, enabling us to see both the deformation of the signal and movement of the tracking copepod. 
 }
\label{fig:motivation}
\end{figure}

Evidence suggests that many organisms  respond to concentration difference (= signal strength) and orient themselves to the desired direction, either locating towards or escaping from a source~\cite{Buck2000,Vickers2000,Johnson2012}. 
%
%
%
%
%
Biological and physical gradients also act as signals for tracking processes in smaller organisms; see, e.g.,~\cite{Woodson2005} and references within.  Copepods, a type of zooplankton about $0.1$ cm in length,  
 are known to aggregate at the boundaries of different water bodies in the ocean~\cite{Holliday1998}. This aggregation is thought to be a result of the response to oceanic structures involving spatial gradients of flow velocities and densities~\cite{Woodson2005}. Copepods adjust their swimming speed or turning frequencies with respect to these physical gradients in the water environment. Also, copepods sense biological gradients in mate-seeking~\cite{Doall1998}.  In careful laboratory experiments by Jeannette Yen that focus on the mating behavior of the copepod {\em Temora longicornis}, a chemically-scented trail that mimics the pheromone-laden trail of the female is introduced into a quiescent water tank. Male copepods are able to detect and successfully track the trail mimic to its source as shown in FIG~\ref{fig:motivation}.


%


In this work, we are loosely inspired by the copepod tracking ability of the female chemical trail. We develop an idealized, simple model where a moving chemical source generates a trail in an \yh{infinite} two dimensional space and a tracker is able to locally sense the chemical field and adjust its orientation accordingly to locate and track the trail. The organization of this work is as follows. We first illustrate the chemical trail in section~\ref{sec:problem} by reformulating the problem in the moving frame attached to the source. 
In section~\ref{sec:tracking}, we study the conditions for successful tracking using a gradient-based tracking scheme. In the situation where the tracker is far away from the chemical trail such that the gradient information is not reliable, a random-walk phase is introduced to first detect the chemical signal before switching to the gradient-tracking method. The detection algorithm and results are described in Section~\ref{sec:detection}. We conclude by summarizing our findings and discussing their potential implications to understanding the behavior of copepods in section~\ref{sec:conclusion}.


%


\section{Problem description}\label{sec:problem}

\begin{figure*}[!t]
\includegraphics[scale=0.9]{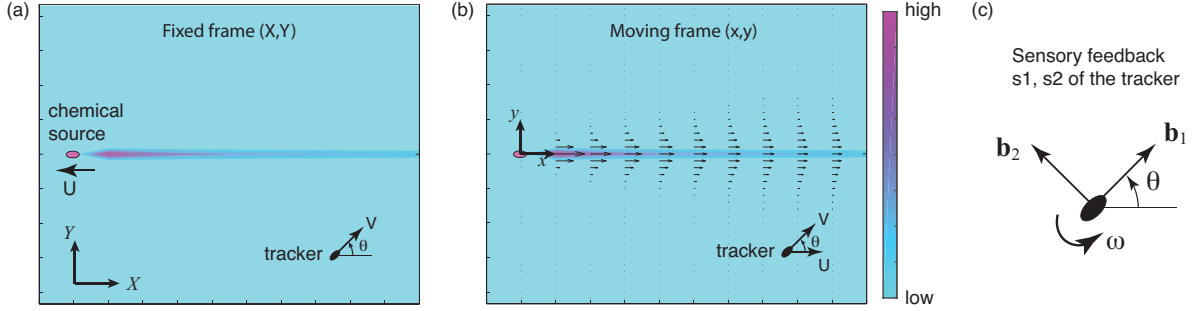}
\caption{Chemical trail left behind a source moving at constant speed $U$ to the left: (a)  in  fixed inertial frame $(X,Y)$ and (b) in a frame $(x,y)$ moving with the source. A sensory system (black ellipse) moving at a constant velocity $V$  senses the chemical gradient in  body-fixed frame $\mathbf{b}_1,\mathbf{b}_2$ and adjusts its orientation $\theta$ accordingly (c). 
}
\label{fig:concentration}
\end{figure*}

Consider a chemical source moving at a constant velocity $U$ from right to left in a fixed frame $(X,Y)$, shown in FIG.~\ref{fig:concentration}(a). The concentration field is governed by the diffusion equation
\begin{equation}
\frac{\partial C}{\partial t} = K\frac{\partial^2 C}{\partial Y^2} + Q\delta(X+Ut)\delta(Y),
\label{eqn:moving}
\end{equation}
where $Q$ is the  rate of  generation of the chemicals, $K$ is the mass diffusivity of the chemicals, and $\delta$ is the Dirac-delta function. In~\eqref{eqn:moving}, we neglected diffusion in the  $X$-direction. This assumption can be readily justified by calculating the P\'{e}clet number Pe, defined as  
 the ratio of advective to diffusive transport rate. Large Pe (Pe $\gg 1$) implies that advection is dominant while for small Pe (Pe $\ll 1$) diffusion is dominant. In the $X$-direction, P\'{e}clet number is given by Pe$ = {LU/K}$ where $L$ and $U$ are the characteristic length and speed, which for a swimming copepod take the values $L = 0.1{\rm cm}$ and $U = 1{\rm cm/s}$~\cite{Woodson2007}.  The
  diffusivity coefficient involved in small biological organisms is of the order $K = 10^{-5} {\rm cm^2/s}$~\cite{Lombard2013}. Thus,   Pe $ \sim 10^4$ and diffusion is negligible  in the $X$-direction. 
  
It is convenient to rewrite equation~\eqref{eqn:moving} in a reference frame $(x,y)$ moving with the chemical source at a speed $U$, shown in FIG. 2(b). The moving frame $(x,y)$ is related to the fixed inertial frame $(X,Y)$ via the transformation
\begin{equation}
x = X+Ut, \quad y = Y
\label{eqn:frame}
\end{equation}
Therefore, $\dfrac{\partial C}{\partial t} = \dfrac{\partial C}{\partial t} + U\dfrac{\partial C}{\partial x}$ and equation~\eqref{eqn:moving} becomes an advection-diffusion equation as
\begin{equation}
\frac{\partial C}{\partial t} + U\frac{\partial C}{\partial x} =K\frac{\partial^2 C}{\partial y^2} + Q\delta(x)\delta(y).
\label{eqn:fixed}
\end{equation} 
The steady-state solution of~\eqref{eqn:fixed} is given by 
\begin{equation}
C = \frac{Q/U}{\sqrt{4\pi (K/U) x}}\exp\left(-\frac{y^2}{4(K/U)x}\right).
\label{eqn:concentration}
\end{equation}
A color map of this concentration  is shown in FIG.~\ref{fig:concentration} with pink indicating higher concentration values. 

 Next we study the tracking behavior of a sensory system, or a chemical tracker, in response to these chemical signals. One natural example of chemical tracking is in the mating behavior of copepods, where the female swims at a roughly constant speed along a straight path, while the male swims at faster speeds along a sinuous route until it detects the chemical trail left by the female and follows it~\cite{Woodson2007}. The female copepod has  body length $L = 0.1 \rm cm$ and speed around $U = 1 {\rm cm/s}$, leaving a trail of chemicals where $Q/U = 0.253 {\rm \mu g/cm^3}$, or equivalently, the source rate is $Q = 0.253 {\rm \mu g/(cm^2\cdot s})$. We inherit these parameter values for our current study. In our simulation, we choose mass scale $m^* = 0.1\mu g$, velocity scale $U^* = U = 1 \rm cm/s $ and length scale $L^* = 10L = 1{\rm cm}$ to non-dimensionalize the problem. This choice of length scale makes it more feasible to treat the tracker as a point particle. 

\section{Tracking of Chemical Trails}\label{sec:tracking}

Consider a sensory system moving at a swimming speed $V$ and
let $\left(\bm{b}_1,\bm{b}_2\right)$ 
be an orthonormal frame attached to the sensory system such that $\mathbf{b}_1$ is aligned along the swimming direction; see FIG.~\ref{fig:concentration}. Let $\theta$ denote the orientation of the $\bm{b}_1$-axis measured from the $\bm{e}_1$ direction. The sensory system is able to sense the directional concentration gradients $s_1 = \nabla C \cdot \mathbf{b}_1$ and $s_2 = \nabla C \cdot \mathbf{b}_2$ and adjust its orientation, but not speed, based on the gradients it senses. In the moving frame $(x,y)$, we have
\begin{equation}
\label{eq:eom}
\dot{x} = U + V\cos\theta, \quad \dot{y} = V\sin\theta, \quad \dot{\theta} = F(s_1,s_2).
\end{equation}
Here, we postulate a simple form of the function  $F(s_1,s_2)$, namely,
\begin{equation}
\label{eq:F}
 F(s_1,s_2) =  \omega \, \mathrm{sgn}(s_2)H(\gamma - s_1)
\end{equation}
where $\omega$ is a constant rotation rate, sgn$(\cdot)$ is the sign function and $H(\cdot)$ is the heaviside function. According to~\eqref{eq:F}, if the concentration gradient $s_1 = \nabla C \cdot \mathbf{b_1}$ in the $\mathbf{b}_1$-direction is larger than a threshold value $\gamma$, then  $\dot{\theta}=0$ and the sensor continues to move in the same direction. If $s_1$ is less than $\gamma$,  one has $H(\gamma-s_1) = 1$ and $\dot{\theta} = \omega \, \mathrm{sgn}(s_2)$. In this case, the sensor turns with angular velocity $\omega$ into the direction of increasing concentration, indicated by the sign of $s_2 = \nabla C \cdot \mathbf{b_2}$. Note that the tracking scheme depends on the sign of the signals instead of their exact values. Therefore, the results are not sensitive to the distance between the tracker and the chemical source, especially in the $x$-direction. 
\begin{figure}[!t]
\includegraphics[scale=0.8]{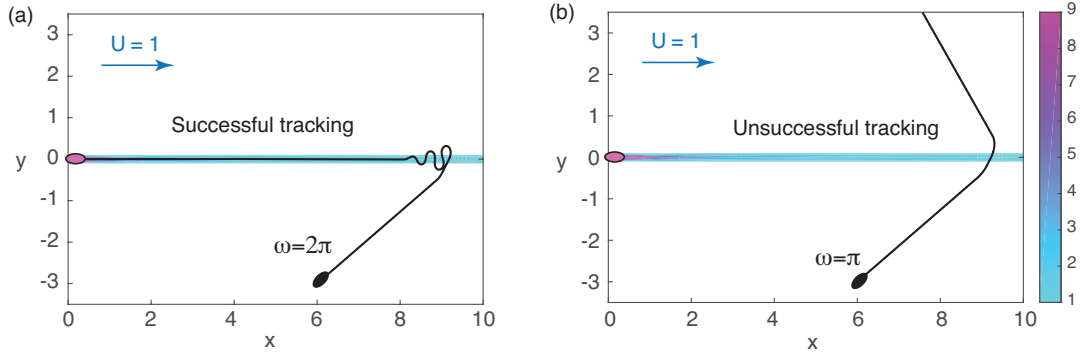}
\caption{Successful and unsuccessful tracking for parameter values (a) $\omega = 2\pi, V = 2$ and (b) $\omega = \pi, V = 2$. Other parameters are set as $Q=0.1$, $K = 10^{-5}$, and $\gamma = 0.01$. The initial location of the sensory system is $x(0) = 6, y(0) = -3$ and its initial orientation is $\theta(0) = \pi/3$. Colors represent the steady-state spatial distribution of the chemicals.}
\label{fig:track1behavior}
\end{figure}

We simulate the trajectory of our sensory system by integrating equation~\eqref{eq:eom} in time using the adapted-time-step function `ode45' in MATLAB. Basic parameter values are chosen as follows: source rate $Q= 0.1$, diffusivity $K = 10^{-5}$ and and threshold $\gamma = 0.1$. The initial location of the sensory system is $x(0) = 6, y(0) = -3$. FIG.~\ref{fig:track1behavior} shows the trajectories for the same initial orientation $\theta(0) = \pi/3$ and swimming speed $V=2$ but two sets of control parameters: (a) $\omega = 2\pi$ and (b) $\omega = \pi$. In (a), the tracker successfully follows the chemical trail while in (b) the tracker encounters the trail but fails to track it. 
This is because its angular velocity $\omega= \pi$ is not large enough for the sensory system to make a quick turn into the chemical trail. It is worth noting that the oscillatory trajectory in successful tracking is also found in the copepod experiments~\cite{Woodson2007}. 

\begin{figure}[!t]
\includegraphics[scale=0.85]{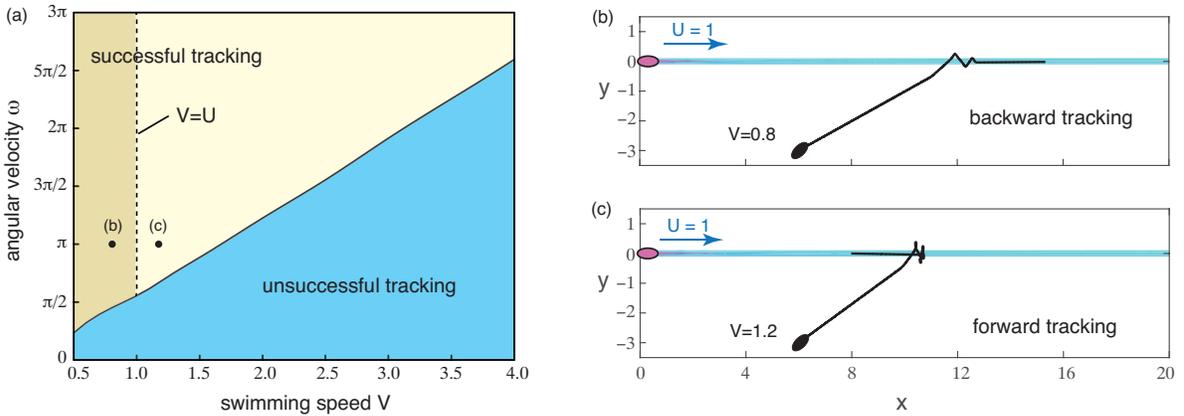}
\caption{ Tracking behavior in the parameter space of swimming speed $V$ and angular velocity $\omega$. Other parameter values and initial conditions are the same as in FIG.~\ref{fig:track1behavior}. (a) Parameter space $(V, \omega)$ of successful versus unsuccessful tracking. (b) Successful backward tracking with swimming speed $V = 0.8$ less than the source speed $U=1$. (c) Successful forward tracking with $V = 1.2 > U$.}
\label{fig:phasespace}
\end{figure}

{We now examine the tracking behavior of the sensory system with respect to the two control parameters: angular velocity $\omega$ and swimming speed $V$. Other parameter values and initial conditions remain the same as those in FIG.~\ref{fig:track1behavior}. We map the unsuccessful and successful tracking on the two dimensional space $(V,\omega)$ in FIG.~\ref{fig:phasespace}(a). It shows that as the tracker swims faster, the required angular velocity $\omega$ for successful tracking also increases. The transition from unsuccessful to successful tracking displays a linear relationship between the angular velocity $\omega$ and the swimming speed $V$. Note that the chemical source has a speed $U = 1$. When the tracker's speed $V < U$, the tracking is in the opposite direction of the source location, which we denote as backward tracking. See the example in FIG.~\ref{fig:phasespace}(b) for $V = 0.8$ and $\omega = \pi$. \yh{Backward tracking of a chemical trail has already been observed in copepod experiments~\cite{Doall1998}.} When the tracker's speed is larger than that of the source, $V>U$, it tracks the chemical trail in the direction towards the location of the source, termed as forward tracking, shown in FIG.~\ref{fig:phasespace}(c) for $V = 1.2$ and $\omega = \pi$. The boundary separating these two types of successful tracking is illustrated as a dashed line at $V=U$. \yh{This boundary can be easily inferred from equation~\eqref{eq:eom} by setting $\theta = \pi$ where the tracker is heading into the direction of the source. To achieve forward tracking, the horizontal velocity $\dot{x}$ must be in the negative $x$-direction; namely $\dot{x} = U-V < 0$.} Both backward and forward tracking are successful in tracking the chemical trail but differ in their ability to locate the source. 

\begin{figure}[!h]
\includegraphics[scale=0.85]{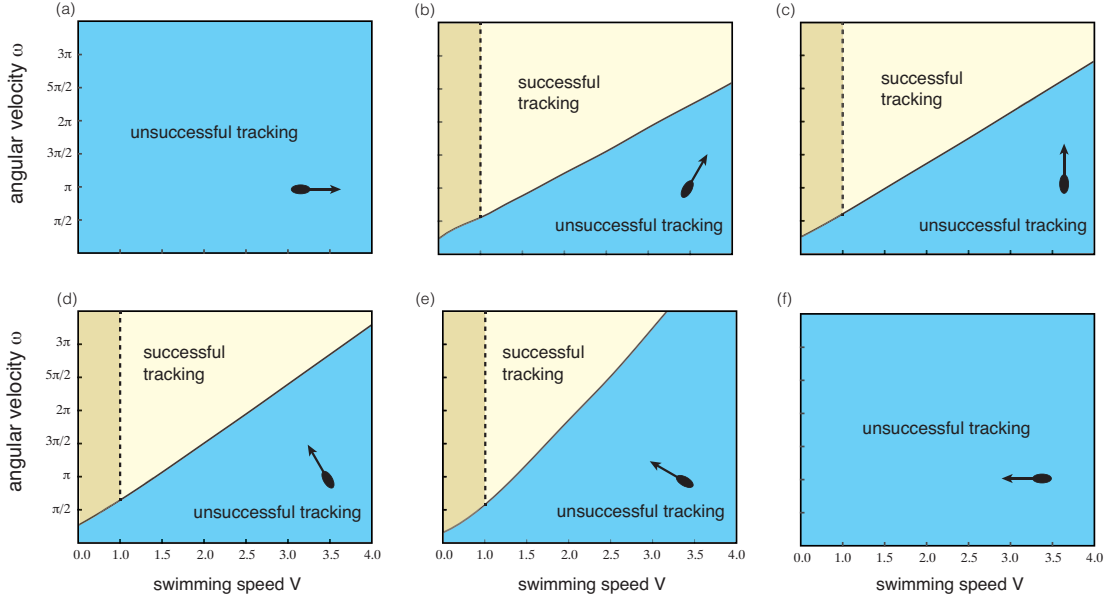}
\caption{Parameter space of  swimming speed $V$ and angular velocity $\omega$ as a function of six different initial orientations  $\theta(0) = 0$, $\pi/3$, $\pi/2$, $2\pi/3$, $5\pi/6$ and $\pi$, shown in (a)-(f) respectively. Dashed line is the interface between backward and forward tracking.}
\label{fig:phasespace2}
\end{figure}

The parameter space displayed in FIG.~\ref{fig:phasespace}(a) is specified for one initial orientation $\theta(0) = \pi/3$. We now explore the two dimensional space $(V,\omega)$ in FIG.~\ref{fig:phasespace2}(a)-(f) with respect to six different initial orientations: $\theta(0) = 0, \ \pi/3, \ \pi/2, \  2\pi/3, \  5\pi/6$ and $\pi$. When $\theta(0) = 0$ or $\pi$, the tracker moves in the positive or negative $x$-direction parallel to the chemical trail without ever turning or intercepting the trail. 
The sensory system fails to approach the chemical trail and therefore the tracking is unsuccessful irrespective of the values of $V$ and $\omega$. For initial conditions that intercept the trail, both unsuccessful and successful tracking can be achieved, as shown in FIG.~\ref{fig:phasespace2}(b)-(e). Note that plot (b) is the same as the one in FIG.~\ref{fig:phasespace}. The boundary marking the transition from unsuccessful to successful tracking is given by a linear relationship between $V$ and $\omega$. 
 The slope of the linear boundary gets steeper as $\theta(0)$ increases. 
In other words, as the angle between the tracker $\theta(0)$ and the trail becomes more obtuse, the tracker requires faster  rotational motion for successful tracking.
As $\theta(0)\to \pi$, the slope of the transition between unsuccessful and successful tracking tends to infinity. In FIG.~\ref{fig:phasespace2}(b)-(e), the transition from  backward to forward tracking is independent of $\theta(0)$ and bifurcates at $V = U$. 

\section{Detection of Chemical Trails} \label{sec:detection}

The gradient-based model for trail tracking is not feasible for the initial detection of the trail because at distances far away from the trail the gradient is too shallow to be accurately sensed. However, the local concentration itself can be sensed~\cite{Li2006ADA}. Therefore, we introduce a detection step to first find the strong chemical trail by comparing the local chemical concentration to a threshold value $C_o$. If the former is larger, then the chemical trail is detected and the tracker enters the tracking step using gradient information in~\eqref{eq:F}. During the detection, the tracker executes a random walk that resembles the run-and-tumble behavior of bacteria~\cite{Adler1966,Berg1972}. That is to say, the tracker runs in the same orientation if the detected concentration is increasing otherwise it tumbles by randomly choosing a direction. An illustration of the detection algorithm is shown in FIG.~\ref{fig:flowdiagram}. 

According to FIG.~\ref{fig:flowdiagram},  a tracker initially at $(x_m, y_m)$ detects the local concentration as $C_m = C(x_m,y_m)$ and picks a random direction $\theta_m$ to start moving. After a given time  $t_m= t_{m-1} + \triangle t$, where $\Delta t$ is the time step and $m$ is a positive integer,  the tracker's position is given by $x_m = x_{m-1} + V \cos (\theta_{m-1}) \Delta t$ and $y_m = y_{m-1} + V \sin (\theta_{m-1}) \Delta t$. It senses a new concentration $C_{m}$ at the new location $(x_m,y_m)$. If $C_{m} > C_o$, then the chemical trail is detected. Otherwise, the tracker executes the run-and-tumble behavior by comparing the current concentration $C_m$ to the previous one $C_{m-1}$. If $C_m > C_{m-1}$, the tracker runs without changing its orientation $\theta_m = \theta_{m-1}$. If $C_m < C_{m-1}$, then the tracker picks a random direction $\theta_m$ and follows that direction for $N$ time steps. \yh{That is, $1/(N\Delta t)$ can be interpreted as the ``frequency" of random walk.} At the time when the local concentration $C_m$ achieves the threshold value $C_o$, the detection step ends and transitions to the gradient-tracking step. The total detection time is calculated as $t_d = m\triangle t$. Yet, if during a simulation, the time count $m$ is greater than a given maximum value $m_{\rm max}$, then the detection is considered unsuccessful within the given amount of time $t_{\rm max} = m_{\rm max}\triangle t$. Note that if time is long enough, the tracker is guaranteed to detect the chemical trail in a two dimensional plane~\cite{Borwein2004,Spitzer2013}. 

\begin{figure}[!t]
\includegraphics[width = \linewidth]{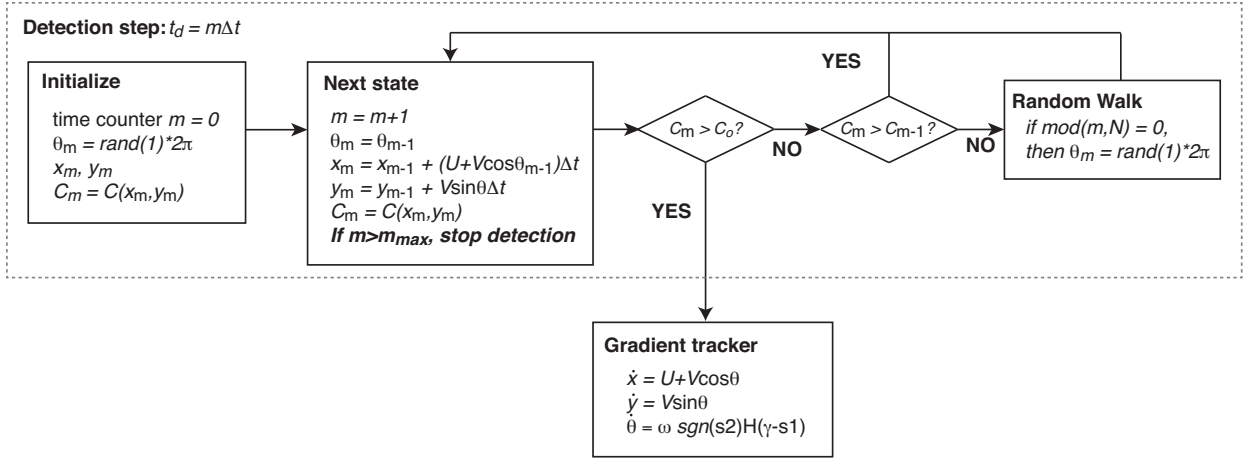}
\caption{ The run-and-tumble detection step of the chemical tracker. $C_o$ denotes the concentration threshold for the chemical trail. The tracker can detect the concentration $C_m$ at its current location $(x_m,y_m)$ and also remembers the previous concentration $C_{m-1}$. If $C_m < C_o$, the tracker runs and tumbles in the detection step.  If $C_m > C_{m-1}$, it remains the current orientation $\theta_m$. Otherwise, it executes a random walk with frequency $N$ such that $\theta_m$ is randomly chosen in between 0 and $2\pi$ every $N$ iterations of time step $\triangle t$. When $C_m > C_o$, the tracker stops the detection step and enters the tracking behavior. If $m > m_{\rm max}$, then the detection is unsuccessful.
}
\label{fig:flowdiagram}
\end{figure}

\begin{figure}[!h]
\includegraphics[scale=0.85]{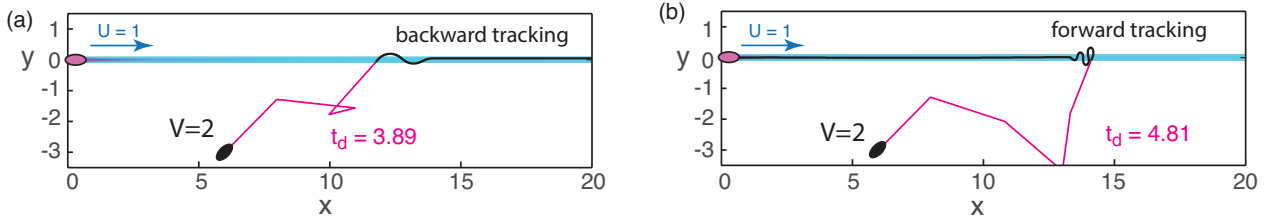}
\caption{Two sample cases of random walk in the detection of chemical trail. Control parameters are set as follows: concentration threshold $C_o = 10^{-5}$, time step $\triangle t = 0.01$, random-walk frequency $N=100$ and maximum detection time $t_{\rm max} = 20$. The two trackers have the same initial conditions $x(0) = 6, y(0) = -3, \theta(0) = \pi/3$ and parameter values $\omega = 2\pi, V = 2$ but end up with different tracking behaviors: (a) backward tracking and (b) forward tracking. }
\label{fig:random}
\end{figure}

The detection step is governed by four control parameters: the concentration threshold $C_o$, time step $\triangle t$, frequency $N$, and maximum detection time $t_{\rm max}$. 
Here, we choose $C_o = 10^{-5}$, $\triangle t = 0.01$, $N=100$ and $t_{\rm max}= 20$ and study the effect of swimming speed $V$ on the detection time $t_d$. 
In FIG.~\ref{fig:randomwalk}, we show two sample trajectories starting at the same initial position $x(0) = 6, y(0) = -3$ and random initial orientation $\theta(0) = \pi/3$ for the same parameter values $\omega = 2\pi$ and $ V = 2$. The two trajectories are distinct owing to the random nature of the search motion such that backward tracking occurs in (a) and forward tracking in (b). The detection time $t_d$, which we define as the total time it takes the sensory system to first detect the trail, is also not the same in the two simulations: $t_d = 3.89$ in (a) and $t_d = 4.81$ in (b). Note that the detection time $t_d$ is independent of the angular velocity $\omega$, which only participates in the tracking behavior, but depends on the swimming speed $V$. \yh{In the copepod experiments~\cite{Doall1998}, the dimensional detection time is up to 10 seconds.}


FIG.~\ref{fig:randomwalk} depicts the histogram or distribution of detection time $t_d$ of successful detections obtained from $1000$ distinct simulations for the same initial locations shown in FIG.~\ref{fig:random} and three parameter values of $V = 1, 2, 3$. We fit the probability distributions to smooth exponential functions $P(t) = \lambda e^{-\lambda t}$, shown as red curves, such that the average detection time is $\left\langle t_d \right \rangle = 1/\lambda$. \yh{Note that the exponential fit is not perfect, but it is a closer analytical fit to the resulting distribution compared to a Poisson and normal distribution. The discrepancy between the analytical fit and the numerical data has minimal implications on the following results.} As velocity increases, the decrease of the probability density function is steeper (larger $\lambda$) and the averaged detection time $\left\langle t_d \right \rangle$ decreases from $9.83$ to $6.39$ and $4.39$. Therefore, larger swimming speed results in faster detection. In addition, out of the 1000 simulations we run, we keep track of the number of simulations which resulted in unsuccessful detection in $t_{\rm max} = 20$ . We find that the ratio of unsuccessful detection to total number of simulations is $0.57, 0.3$ and $0.26$ for $V = 1, 2$ and $3$, respectively. That is to say, faster swimming is also beneficial to more successful detections in a given amount of time. 

\begin{figure}[!h]
\includegraphics[scale=0.85]{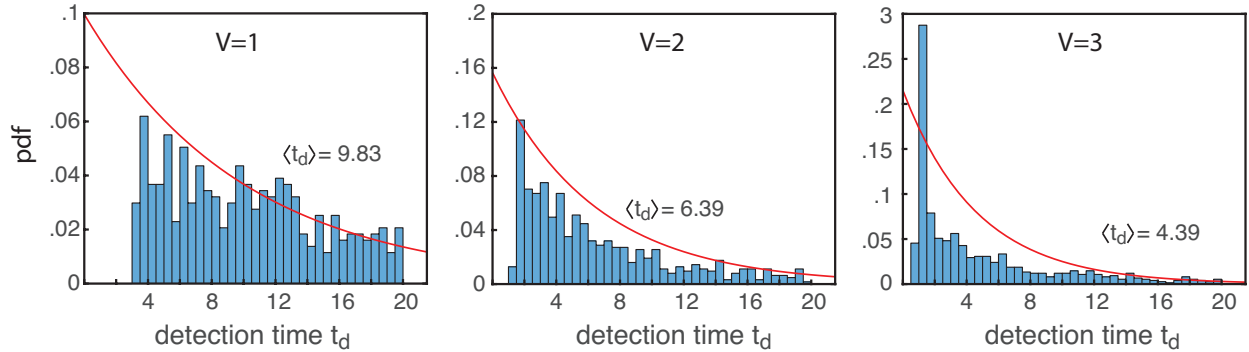}
\caption{Probability distribution of detection time $t_d$ with respect to different swimming speed $V$. The red curves are the fitted exponential probability density functions $P(t) = \lambda e^{-\lambda t}$ with $\lambda = 1/\left\langle t_d \right\rangle$, where $\langle t_d \rangle$ is the averaged detection time. For each value of $V$, 1000 distinct simulations are conducted to obtain the distribution of successful detections. The values of control parameters and also the initial locations of the tracker are the same as those in FIG.~\ref{fig:random}.}
\label{fig:randomwalk}
\end{figure}

We finally evaluate the average detection time $\langle t_d \rangle$ as a function of the tracker's initial location $x(0), y(0)$. The region of interest is chosen to be $[0.1,10] \times [-5,-1]$ as shown in FIG.~\ref{fig:randomphase}. The colors indicate the values of the detection time at the corresponding initial locations, with red specifying longer detection time. We can see that $\langle t_d \rangle$ varies little in the $x$-direction especially in the range of $x(0)>2$ meanwhile the average detection time decrease significantly when the horizontal distance between the tracker and the source is small $x(0)<1$. The average detection time grows with increasing distances in the $y$-direction. These finding are consistent with the intuition that closer distance between the tracker and the source results in faster detection.

\begin{figure}[!t]
\includegraphics[scale=0.85]{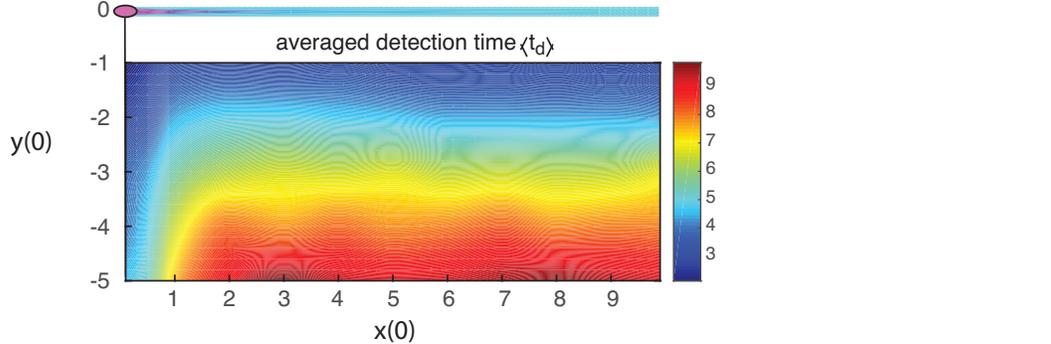}
\caption{ The averaged detection time $\langle t_d \rangle$ as a function of the tracker's initial location $x(0),y(0)$ in the region of $[0.1,10] \times [-5,-1]$. Red colors represent longer detection time.}
\label{fig:randomphase}
\end{figure}

\section{Conclusions} \label{sec:conclusion}


Odor tracking plays an important role in the behavior of organisms at different scales and in different environments and could have significant implications on engineering and robotic applications. Inspired by the odor-tracking abilities of male copepods in their mating behavior, we simulated a sensory system that tracks and detects a two-dimensional chemical trail generated by a moving chemical source. The tracker can sense the local chemical gradients in its own body-fixed frame. The sensed gradients are used to control the orientation of the tracker such that it turns into the direction of increasing concentration. 
We identify the tracking behavior as successful if the trajectory of the tracker ends up oscillating around or directly moving inside the chemical trail. Otherwise, the tracking is unsuccessful. Successful tracking consists of either backward or forward tracking. If the sensory system successfully tracks the trail but moves away from the source, then it is backward tracking. Backward tracking occurs when the speed of the tracker is less than that of the chemical source. 

We then mapped the tracking behavior onto the parameter space consisting of the speed $V$ of the tracker normalized by the speed of the chemical source and the angular velocity $\omega$ of the tracker. The results show that higher $V$  requires larger $\omega$ for successful tracking. The boundary marking the transition from unsuccessful to successful tracking follows a linear growth of $\omega$ as a function of increasing $V$.  The parameter space $(V, \omega)$ changes with respect to the initial orientation $\theta(0)$ such that when the angle between the tracker and the trail becomes more obtuse (i.e., the angle between the velocity of the tracker and the velocity of the source is more shallow)  the tracker requires both larger speed and angular velocity to succeed in tracking. That is to say, the tracker should speed up to successfully track the trail when the orientation of its velocity is close to that of the source. 

A detection step is introduced when the chemical gradient is too weak to be accurately sensed by the tracker such as when the tracker is located far from the chemical trail. In this situation, the sensory system detects the chemical concentration first until the sensed local concentration is larger than a threshold value. The detection step is adapted from the run-and-tumble behavior of bacteria such as \textit{E.~coli}, which runs when sensing a chemical signal or tumbles otherwise. In our implementation, the tracker continues in the same orientation if it senses an increasing concentration in that direction. If not, the tracker randomly picks an orientation $\theta$ from $0$ to $2\pi$. If the detection takes longer than a given amount of time (the total simulation point), then the detection is unsuccessful. We illustrated the distribution of the detection time $t_d$ obtained from 1000 distinct simulations and calculated the average detection time of successful ones. We found that the average detection time decreases with increasing speed $V$ and the ratio of successful detection is higher when $V$ is larger. Therefore, for a more successful detection, a fast speed $V$ is preferred. We also showed that closer initial location to the source results in smaller detection time.


The two main results obtained from this study -- the fact that both successful detection and successful forward tracking require the tracker to have larger swimming speed than the source and that the tracker's speed should be even larger when it swims nearly parallel to the source -- are consistent with experimental observation of the copepod mating behavior~\cite{Doall1998}. Male copepods are known to swim faster than female copepods. While the reasons may be biological, this difference in speed between the male and female seems to have significant implications on successful detection and tracking of the female. 
\yh{Further, the detection time scale obtained here is consistent with experimental measurements of copepods~\cite{Doall1998}. Male copepods are reported to detect the chemical trail in time intervals up to 10 s, which is similar to the average detection time reported in this study. Note that here one dimensionless unit time scales to 1s. }

A few remarks on the limitations of the model and future directions are in order. We considered a simple gradient-tracking model where the speed of the tracker and its turning rate are not affected by the intensity of the chemical signal. While this model was able to track the chemical trail, it would be interesting in future studies to compare  this model to more complex models where the speed and turning rate to change with the chemical signal. This study was restricted to two-dimensional tracking and detection but in many aquatic organisms, this behavior is inherently three-dimensional. Also, we considered the chemical signal to diffuse in a quiescent environment. In many real-world applications, the environment is often characterized by unsteady and at time turbulent flows. Future work will extend the framework presented here to account for three-dimensional effects and the effect of flows and patchiness in the chemical signal~\cite{Weissburg1994,Kennedy1974,Ishida1996,Kanzaki1996,Belanger1998}. It is also interesting to couple the sensory and control framework presented here to more accurate models of the swimming mechanics;  see~\cite{Alben1545} for an analysis of the details of such drag-based swimming and~\cite{Catton2007} for an experimental study of the flow field generated by the swimming motion. 




Acknowledgment. This work is partially supported by  the Office of Naval Research through the grant ONR
14-001.

\bibliography{Copepod}

\end{document}